\documentclass[a4,12pt]{article}
\usepackage[dvips]{graphicx}
\usepackage{amsmath,amssymb}
\def\be{\begin{equation}}
\def\ee{\end{equation}}
\newcommand{\Mat}[1]{{{\boldsymbol{#1}}}}

\date{}
\begin{document}

\title{\bf Scalar ether theory of gravity: \\
a modification that seems needed}

\author{Mayeul Arminjon}
\maketitle 
%
\begin{center} 
Laboratoire ``Sols, Solides, Structures'' \\
(Unit\'e Mixte de Recherche of the CNRS), 
\\ BP 53, F-38041 Grenoble cedex 9, France

\vspace{1.5cm}

\begin{abstract}
The construction of the scalar theory based on the concept of gravity as Archimedes' thrust is briefly summarized, emphasizing the two (extreme) possibilities that result from this concept for the gravitational rod contraction: it can either occur in only one direction, or be isotropic. A modified equation for the scalar field is stated for the new, isotropic case. The reasons to consider this case are: i) it is almost as natural as the other case, and ii) it should avoid the violation of the weak equivalence principle, found for a small extended body with the directional contraction. The dynamical equation stays unchanged. 
\end{abstract}
\end{center}

\newpage

\section{Gravity as Archimedes' thrust}
A mechanism for gravity was proposed by the author, which turns out to be quite similar to a mechanism proposed by Euler \cite{Euler} to interpret Newton's gravity. In both mechanisms, the elementary contituents of matter must have a finite extension and must have all the same mass density. Gravity may then be interpreted as Archimedes' thrust in a fluid ``ether" surrounding any elementary contituent. In the author's mechanism \cite{A8}, the elementary contituents themselves are seen as localized flows in that very fluid---this is the ``constitutive ether", which allows one to get a picture of the creation/ annihilation/ transmutation of particles, and of the instable ``resonances", all being observed in particle physics. Moreover,  gravitation is assumed to involve only the ``smoothed-out", {\it macroscopic} pressure force. This concept of gravity as Archimedes' thrust due to the macroscopic pressure in the constitutive ether leads to a definite expression for the gravity acceleration:
\be \label{g_rho_e}
\mathbf{g}=-\frac{\mathrm{grad}p_e}{\rho_e},
\ee
where $p_e$ is the (macroscopic) ``ether pressure" and $\rho_e = \rho_e(p_e)$ is the (macroscopic) ``density" in the ether, the latter being assumed a barotropic fluid. This expression is the starting point of a theory of gravity based on just the scalar field $p_e$. It is natural to assume that Newton's gravity, which propagates instantaneously, should correspond to the limiting case of an incompressible ether ($\rho_e=\mathrm{Constant}$). A compressible ether should then lead to waves of the ether pressure $p_e$, i.e. to gravitational waves.\\

\section{Relativity, clocks and rods}
Special relativity (SR), in the Lorentz-Poincar\'e version, is compatible with the ether \cite{Prokhovnik67}--\cite{Brandes01}. However, in the investigated theory of gravity, the latter is seen as a gradient of pressure and density in the universal fluid, hence it is intrinsically a result of the heterogeneity of ``space" (i.e., in the language of the theory, the heterogeneity of the pressure or equivalently of the density in the constitutive ether)---whereas SR assumes that space is homogeneous. It follows that SR can apply only locally in the presence of gravitation. \\

As is known to participants of this Series of meetings, the Lorentz-Poincar\'e version of SR makes all of SR result from the Lorentz rod contraction and the Larmor clock slowing \cite{Prokhovnik67}--\cite{Brandes01}---the clock slowing being in fact a consequence of the Lorentz contraction and of the ``negative" result of the Michelson-Morley experiment \cite{Prokhovnik67,A9}. Now these two metrical effects of Lorentz-Poincar\'e SR can be interpreted as resulting from a variation in the ``apparent" ether density; hence, in the framework of the investigated model of gravity, one is naturally led to postulate gravitational rod contraction and clock slowing \cite{A9}. The very notion of a clock slowing and a contraction of objects implies that there must be a {\it reference} with respect to which these effects do occur, and the simplest possibility is clearly to assume a flat reference metric. Thus, space-time is endowed with two metrics: a flat ``background" metric $\Mat{\gamma}^0$ and a curved ``physical" metric $\Mat{\gamma}$.\\

However, the rod contraction may either occur in just {\it one direction} (as is the case for the Lorentz contraction), which must then be that of the gravity acceleration, i.e. that of the pressure gradient---or, alternatively, the rod contraction may be assumed to be {\it isotropic}. (One may also invent intermediate solutions.) In Ref. \cite{A9}, and until recently \cite{A35}, the author has always assumed the first, anisotropic possibility, for which the correspondence between metrical effects of motion and gravitation is the closest. The second, isotropic possibility for the gravitational rod contraction, is what Podlaha \& Sj\"odin \cite{PodlahaSjodin} had in mind in their analysis of that correspondence---as it can be checked in Ref. {\cite{Podlaha00}. Their analysis of the correspondence has some similarity with the one proposed in Ref. \cite{A9}, but it differs from the latter even apart from the difference regarding the isotropic or anisotropic rod contraction (see Ref. \cite{B13} for a comparison).\\

\section{Dynamics}
Dynamics of a test particle is based on a non-trivial extension of Planck's special-relativistic form of Newton's second law, involving the gravity acceleration (\ref{g_rho_e}) \cite{A15}. It is then extended to any continuous medium \cite{A20}, including classical fields such as Maxwell's field \cite{B13}. One gets the following preferred-frame equation for the energy-momentum tensor ${\bf T} = (T^{\mu \nu }) $:
\begin{equation} \label{continuum}
T_{\mu;\nu}^{\nu} = b_{\mu},	
\ee
with
\be \label{b_mu}
b_0(\mathbf{T}) \equiv \frac{1}{2}\,g_{jk,0}\,T^{jk}, \quad b_i(\mathbf{T}) \equiv -\frac{1}{2}\,g_{ik,0}\,T^{0k},	          
\ee
where $\Mat{g}=(g_{ij})$ is the space metric in the preferred reference frame of the theory. (Indices are raised and lowered with the space-time metric $\Mat{\gamma}=(\gamma_{\mu \nu})$, and semicolon means covariant differentiation using the Christoffel connection associated with metric $\Mat{\gamma}$.) {\it This equation is valid independently of the possible form for metric} $\Mat{\gamma}$, in particular independently of whether the gravitational rod contraction is isotropic or not.\\

On the other hand, the equation for the scalar gravitational field (the ether pressure $p_e$) is only partly constrained \cite{A35} by the assumption \cite{A8} that Newton's theory has to be recovered for $\rho_e=\mathrm{Constant}$, and that the compressible case should be associated with a wave equation. A further constraint on the scalar field equation is that this equation, together with Eq.(\ref{continuum}), should imply an {\it exact local conservation law for the energy} \cite{A15}, i.e., for that matter, an equation that may be set in the form 
\be \label{localconservation}
\partial_T w_\mathrm{m} + \mathrm{div}(w_\mathrm{m} \mathbf{u})+ \partial_T w_\mathrm{g} + \mathrm{div}(\Mat{\Phi} _\mathrm{g}) = 0,
\ee
where $w_\mathrm{m}$ and $w_\mathrm{g}$ are the material and gravitational energy densities, $\mathbf{u}$ is the material velocity, and $\Mat{\Phi} _\mathrm{g}$ is the gravitational energy flux. With this further constraint, the scalar field equation is closer to being fixed. E.g., in the case with the anisotropic rod contraction, a definite equation has been assumed for $p_e$ \cite{A9,A15}, which thus made the theory entirely closed.\\

\section{Observational test}
Much work has been done to test the theory obtained with the anisotropic gravitational rod contraction. That theory seems a priori promising because it predicts exactly the Schwarzschild exterior metric in the static case with spherical symmetry \cite{A9}. And since it can be proved that no preferred-frame effect does occur for the gravitational effects on light rays, one indeed obtains the same predictions for those effects as the ones deduced in general relativity (GR) from the standard Schwarzschild metric \cite{A19}. The theory has been tested also in celestial mechanics, and it has been found that here too the preferred-frame effects are not a problem (although, in the case of celestial mechanics, there {\it are} indeed such effects according to that theory) \cite{O1}. A similar formula as that used in GR has been derived for the energy loss by gravitational radiation, so the theory should pass the binary-pulsar test \cite{A34}. Cosmology has also been investigated; in particular, the theory predicts that the cosmic expansion must be accelerated, and there is no singularity with infinite density \cite{A28}.\\

However, that version of the scalar ether theory leads to a disturbing violation of the weak equivalence principle for an {\it extended body} considered at the {\it point particle limit} \cite{A33}. This violation is precisely due to the anisotropy of the metric, therefore it was argued that {\it it should also occur in GR in some gauges} \cite{A33}. But it should not occur with the isotropic rod contraction. 
\\

\section{Modified equations}

Therefore, the author is currently investigating in detail the case with the isotropic gravitational rod contraction. In coordinates $(x^\mu)$ that are adapted to the preferred reference frame and Galilean for the flat metric, the curved space-time metric has then Ni's form \cite{Ni72}:
\be \label{metric}
(\gamma_{\mu \nu})= \mathrm{diag}(\varphi ^{-1},-\varphi ,-\varphi ,-\varphi), 
\ee
where, for the ether theory,
\be
\quad \varphi \equiv \left(\frac{\rho_e^\infty}{\rho_e}\right)^2, \quad \rho_e^\infty(T) \equiv \mathrm{Sup}_{\mathbf{x}\in\mathrm{[space]}}\rho_e(\mathbf{x},T). 
\ee
(The coordinates $(x^\mu)$ allow to define the ``absolute time" $T$, such that $x^0 \equiv cT$ with $c$ the velocity of light.) Specifically, one is naturally led, in the framework summarized in Sects. 1 to 3, to set 
\be
\varphi = e^{2\psi }
\ee
while stating for the scalar field $\psi$ an equation based on the flat d'Alembert operator:
\be \label{dalembert}
\square \psi  \equiv \psi _{,0,0}-\Delta \psi  = \frac{4\pi G}{c^2} \sigma, \qquad \sigma \equiv T^{00}.				  
\ee
($T^{00}$ is the (0 0) component of the energy-momentum tensor of matter and nongravitational fields, in any coordinates $(y^\mu)$ that are bound to the preferred reference frame and such that $y^0=cT=x^0$ where $T$ is the absolute time.) Together with the dynamical equation (\ref{continuum}) and with the metric (\ref{metric}), this closes the new version of the theory. In particular, it does lead to an exact local conservation law for the energy. Moreover, the gravitational effects on light rays are the same as those deduced from the ``standard post-Newtonian metric" of general relativity, which match very well with the observations \cite{Will}. Finally, the author has already checked that this modified version of the theory does avoid the WEP violation which was found with the former version based on the same dynamics (\ref{continuum}), but which assumed an anisotropic rod contraction and an equation differing from (\ref{dalembert}) for the scalar field.
\\

\end{document}